\documentstyle{europhys}
\def\And{{\rm and\ }}

\newif\ifboo \boofalse

\def\Review#1{\boofalse{\it #1},}
\def\Name#1{{\sc #1},}
\def\Vol#1{\ifboo Vol. {\bf #1}\else{\bf #1}\fi}
\def\Year#1{\ifboo #1\else(#1)\fi}
\def\Book#1{\bootrue{\it #1},}
\def\Page#1{\ifboo {\rm p. #1}\else{\rm #1}\fi}
\newcommand{\beq}{\begin{equation}}
\newcommand{\feq}[1]{\label{#1} \end{equation}}
\newcommand{\beqr}{\begin{eqnarray}}
\newcommand{\feqr}{\end{eqnarray}}
\def\non{\nonumber}
\def\noi{\noindent}
\begin{document}
\euro{}{}{}{}
\Date{}
\shorttitle{AGAPITOS HATZINIKITAS CLASSICAL AND QUANTUM MOTION ON THE ORBIFOLD 
$\cdots$} 
\title{Classical and quantum motion on the orbifold limit of the 
Eguchi-Hanson metric}
\author{A. Hatzinikitas\inst{1}
\footnote{e-mail:hatzinik@insti.physics.sunysb.edu}}

\institute{\inst{1} Institute for Theoretical Physics\\
State University of New York at Stony Brook\\
Stony Brook, NY 11794-3840, USA}

\rec{}{}

\pacs{
\Pacs{12}{60j}{Supersymmetric models}
\Pacs{03}{20}{Classical mechanics of discrete systems}
\Pacs{03}{65}{Quantum mechanics}
      }
\maketitle

\begin{abstract}
We investigate the behaviour of a particle moving on the orbifold limit 
of the EH metric as the two centers approach each other. In the classical 
region of the configuration space 
we specify the physically acceptable solutions and observe a tendency of 
the radial wave function to concentrate around the conical singularity for
small values of its argument. In the quantum case, using Schr\"{o}dinger's 
equation, we determine the energy spectra and the radial eigenfunctions for 
a class of potentials. 
\end{abstract}

%%%%%%%%%%%%%%%%%%%%%%%%%%%%%%%%%%%%%%%%%%%%%%%%%%%%%%%%%%%%%%%%%%%%%%%%%%%%%%
\section{Introduction}

\setcounter{footnote}{0}

Hyper-K\"{a}hler manifolds \footnote{A Riemannian manifold $(M,g)$ 
is hyper-K\"{a}hler if it is equipped with three automorphisms 
$J_{i}; i=1,2,3$ of the tangent bundle which satisfy the quaternion algebra
: $J^{i}J^{j}=-\delta^{ij}+\epsilon^{ijk}J^{k}$, 
$[J^{i},J^{j}]=2\epsilon^{ijk}J^{k}$ and are covariant constant with 
respect to the Levi-Civita connection: $\nabla J^{i}=0$ for $i=1,2,3$.} 
in four dimensions have been studied extensively in connection with the 
theory of gravitational instantons \cite{steven,gilkey}, their algebraic 
generalizations through Penrose's nonlinear graviton theory \cite{penrose} 
and the heavenly equations \cite{tod} over the past twenty years. 

Motivated by their presence in supersymmetric field theories we extract the 
EH metric (which is a member of the complete \footnote{A complete 
hyper-K\"{a}hler manifold is a self-dual, positive definite solution of 
Einstein's equations in vacuum (a self-dual gravitational instanton).} 
regular $SO(3)$-invariant hyper-K\"{a}hler family in four dimensions) from 
the K\"{a}hler potential of the most general $N=2$ nonlinear $\sigma$-models 
action \cite{martin}. Using well known transformations we bring EH metric in 
a more familiar form appropriate for performing our calculations. When the two 
centers of the EH metric coincide the manifold exhibits a singular behaviour 
at $r=0$ and is recognized to be the orbifold $M=C^2/Z_2$. The generalization 
of this statement to all members of the $A_{k}$-series translates into the 
appearance of the manifold $M=C^2/Z_{k+1}$. Taking advantage of the K\"{a}hler 
structure of the manifold we specify an invariant quantity, quadratic w.r.t. 
the Weyl tensor, for which after integration over $SO(3)$ produces the 
correct Hirzebruch signature. The Euler characteristic and the Hirzebruch 
signature are also evaluated for our case study. 

Starting from the classical wave equation describing the motion of a massive 
particle we show that there exist radial polynomial solutions (generalized 
Bessel's functions) which strongly depend on the dimension of our manifold. 
Increasing values of $l$ for $\rho \ll 1$ force the radial function to 
accumulate around the point of highest symmetry (the fixed point $r=0$). The 
angular part is expressed in terms of the Jacobi polynomials which are 
closely related to the associated Legendre functions of the first kind upon 
angle coordinate reduction of the original four dimensional manifold down to 
the three dimensional space. The solutions are suitably normalized in such 
a way that one can recover known orthonormality results in the reduced space.

The quantum behaviour on the manifold is exploited by a detailed study of 
some finite dimensional quantum models. This is achieved with the help of 
Schr\"{o}dinger's equation and switching on different potentials. For 
simplicity we consider two types of potentials (the harmonic  and the 
Coulomb potential) which lead to discretized energy spectra. The energy 
levels of the harmonic potential in one dimension and in the absence of 
angular momenta are identical to the expected ones. In the Coulomb case the 
radial eigenfunctions turn out to be Laguerre's polynomials which reduce to 
the predicted ones in three dimensions.

%%%%%%%%%%%%%%%%%%%%%%%%%%%%%%%%%%%%%%%%%%%%%%%%%%%%%%%%%%%%%%%%%%%%%%%%%%%%%%%%%%%
\section{Rederivation of the Eguchi-Hanson metric}

The N=2 action we consider is \cite{martin}:

\beq
I=\int d^{8}z \left(\sum_{i=1}^{n} \mid\Phi^{i}_{+}\mid^{2} e^{V} 
+ \sum_{i=1}^{n} \mid\Phi_{i-}\mid^{2} e^{-V}-cV \right) 
+ \int d^{6}z \left[ (\sum_{i=1}^{n}\Phi_{i-}\Phi^{i}_{+}-b)S+h.c \right].
\feq{act}

\noi where $d^n z=d^4 x ~ d^{\frac{n-4}{2}} \theta ~ d^{\frac{n-4}{2}} 
\bar{\theta}$, V is a real ($N=1$ four-dimensional) superfield, 
$\Phi_{i\pm}$'s are 2n independent complex chiral (right-, left-handed) 
superfields satisfying
$D_{\alpha} \Phi_{+}=\bar{D}_{\dot{\alpha}} \Phi_{-}=0$, with 
$D_{\alpha}$, $\bar{D}_{\dot{\alpha}}$ the
usual two component spinor covariant derivatives. $\Phi_{i\pm}$'s
can be chosen to be in a representation of $SU(n)$ such that (\ref{act})
is $SU(n)$ invariant. S is a Lagrange multiplier chiral superfield  
satisfying $\bar{D}_{\dot{\alpha}} S=D_{\alpha} \bar{S}=0$, 
the bar and the symbol h.c. denote complex and Hermitian conjugation.
The constants c and b 
are real and complex respectively and parametrize the particular linear 
combination of the coordinate systems we are using to describe the manifold. 
The term $c\int d^8 z V(x,\theta,\bar{\theta})$ represents the 
Fayet-Iliopoulos term which is gauge 
invariant and supersymmetric. The equation of motion for the auxiliary 
fields V and S is:

\beq
\sum_{i=1}^{n} \mid\Phi^{i}_{+}\mid^{2} e^{V} - 
\sum_{i=1}^{n} \mid\Phi_{i-}\mid^{2} e^{-V}=c
\feq{eqm}

\beq
\sum_{i=1}^{n}\Phi_{i-}\Phi^{i}_{+}=b.
\feq{gaug}

\noi We restrict our attention to the case where $b=0$ and there are only  
two chiral complex superfields ($n=2$). Choosing the gauge: 

\beq
\Phi_{+} (x,\bar{\theta})=(\Phi^{1}_{+},\Phi^{2}_{+})=(Z_{1},Z_{2}), 
\Phi_{-} (x, \theta)=(\Phi_{1-},\Phi_{2-})=(-Z_{2},Z_{1})
\feq{gfix}
 
\noi we solve the equation of motion for V ($\sinh V=\frac{c}{2r}$). 
The result is :

\beq
V=\ln\left[\frac{c}{2r}+\sqrt{1+\frac{c^2}{4r^2}}\right]
\feq{eqv}

\noi where $r=\mid Z_{1}\mid^2+\mid Z_{2}\mid^2=Z_{i} Z_{\bar{i}}$ and we 
keep only the positive root of the quadratic equation for V. 
The K\"{a}hler potential is found to be: 

\beqr
K(Z_i,Z_{\bar{i}}) &=& 2r\cosh V-cV \non \\ 
&=& 2r\sqrt{1+\frac{c^2}{4r^2}}-c\ln \left(\frac{c}{2r}
+\sqrt{1+\frac{c^2}{4r^2}}\right)
\label{kahp}
\feqr

\noi and the action now takes the form:

\beq
I=\int d^4 x d^2 \theta d^2 \bar{\theta} K(Z_i,Z_{\bar{i}}).
\feq{lagkah}

\noi The K\"{a}hler metric is given as usual by:

\beq
g_{i\bar{j}}=\frac{\partial^{2}K(Z,\bar{Z})}
{\partial Z^{i} \partial \bar{Z}^j}=
2\delta_{ij}\sqrt{1+\frac{c^{2}}{4r^2}}- 
Z_{\bar{i}}Z_{j}\frac{c^2}{2r^3\sqrt{1+\frac{c^2}{4r^2}}}
\feq{kahm}

\noi with inverse:

\beq
g^{i\bar{j}}=(g_{i\bar{j}})^{-1}=\frac{1}{2}\delta^{ij}
\sqrt{1+\frac{c^2}{4r^2}}-\frac{c^2}{8r^3} 
\frac{\epsilon^{ik}\epsilon^{\bar{j}\bar{m}}  
Z^{k} Z^{\bar{m}}} {\sqrt{1+\frac{c^2}{4r^2}}}.
\feq{ingk}

\noi  We recognize this metric to be the Calabi metric which in four real 
dimensions is the Eguchi-Hanson metric, {\it i.e.\/}, the simplest 
Asymptotically Locally Euclidean (ALE) gravitational instanton 
\footnote{ALE spaces describe a Riemannian 4-manifold geodesically comlete 
and such that: 

\begin {enumerate} 

\item the curvature 2-form is anti(self)- dual 

\item the Riemannian metric is required to approximate a Euclidean 
metric up to $O(r^{-4})$, $g^{ij}=\delta^{ij}+a^{ij}=\delta^{ij}+O(r^{-4})$ 
with appropriate decay in the derivatives of $g^{ij}$. In other words, 
$\partial^{p} a^{ij}=O(r^{-4-p})$, $p\geq 0$ where $r^2=\sum x^{2}_{i}$ and 
$\partial$ denotes differentation w.r.t. the coordinates $x_i$.

\end{enumerate} 

This would agree with the intuitive picture of instantons 
as being localized in finite regions of space-time. The above picture 
is only verified modulo an additional subtlety: the basic manifold at 
infinity resembles a quotient $R^4/\Gamma$, $\Gamma$ being a finite group 
of identifications.}. To prove this we change to Cartesian coordinates
\footnote{From now on we treat the components ($Z_i$) of the
chiral superfields ($\Phi^{i}_{\pm}$) as ordinary coordinates on the
four-dimensional space.} 
using the transformations (the Jacobian of the transformation is 
$(r\sin\theta) /16$):

\beq
Z^{1}=\sqrt{r} \cos(\frac{\theta}{2})e^{\frac{i}{2}(\psi + \phi)},
Z^{2}=\sqrt{r}\sin(\frac{\theta}{2})e^{\frac{i}{2}(\psi - \phi)}
\feq{egha}

\noi and setting $\alpha^{2}=\frac{c^2}{4}$ one finds for  the line element:

\beq
ds^{2}_{EH}=g_{i\bar{j}}dZ^{i}dZ^{\bar{j}}=\frac{1}{4}
\frac{dr^2}{r\sqrt{1+(\frac{\alpha}{r})^2}} 
+\frac{r}{\sqrt{1+(\frac{\alpha}{r})^2}}\sigma^{2}_{z} 
+ r \sqrt{1+(\frac{\alpha}{r})^2}(\sigma^{2}_{x}+\sigma^{2}_{y})
\feq{ehan}

\noi where $\sigma_{x}$, $\sigma_{y}$ and $\sigma_{z}$ are the 
left-invariant one-forms of the $SU(2)$ group (see Appendix A for 
their definition). The variables $r, \theta, \psi, \phi$ are constrained 
by $\alpha\leq r \leq \infty$, $0\leq \theta \leq \pi$, 
$0\leq \phi \leq 2\pi$, $0\leq \psi \leq 2\pi$. The restricted range of 
$\psi$ reflects the $Z_{2}$ identification of antipodal points. 
Using the coordinate transformation $r^2=\rho^4$ and that $\alpha^2=a^4$ 
we end up with the familiar form of the Eguchi-Hanson metric \cite{eguchi}:

\beq
d\tilde{s}^{2}_{EH}=\frac{1}{\sqrt{1+(\frac{a}{\rho})^4}}
(d\rho^2+\rho^2\sigma^{2}_{z}) + \rho^2 
\sqrt{1+(\frac{a}{\rho})^4}(\sigma^{2}_{x}+\sigma^{2}_{y}).
\feq{ehann}

\noi The determinant of the metric in this case is given by $g=\frac{1}{64} 
\rho^6 \left[1+(\frac{a}{\rho})^4\right]^{-1} \left[1
+(\frac{a}{\rho})^4 \sin^2 \theta\right]$.

We can also write the line element in the equivalent form:

\beq
ds^2=V^{-1}(d\phi+ A_{i}dx^{i})^2+V \gamma_{ij}dx^{i}dx^{j},
\feq{alter}

\noi where $V, A_{i}, \gamma_{ij}$ being all independent of $\phi$, and $V$ 
satisfies:

\beq
V(\vec{x})=\epsilon+2m \sum_{i=1}^{k+1}\frac{1}{\left|\vec{x}
-\vec{x}_i \right|}
\feq{vri} 

\beqr
\vec{\nabla} V\!\!&=&\!\! -\vec{\nabla} \times \vec{A} \\
\vec{\nabla} \cdot \vec{\nabla}V \!\!&=&\!\! 2m\sum_{i=1}^{k+1}
\delta^{(3)}(\vec{x}-\vec{x}_i).
\label{curl}
\feqr

\noi The choice $\epsilon=0$, $m=\frac{1}{2}$ corresponds to an admissible 
metric on the $k^{th}$ representative of the $A$-series space. For the EH 
instanton ($k=1$) after some straightforward calculations we find that: 
\footnote{In \cite{olivier} the coefficient $\frac{1}{4}$ of equation (32) 
should be replaced by $\frac{V}{4}$.}

\beqr
V^{-1} & = & \frac{1}{4}(c_{1}\sin^{2}\theta \cos^{2}\psi 
+c_{2}\sin^{2}\theta\sin^{2}\psi+ c_{3}\cos^{2}\theta) \\
\vec{A} & = & \frac{1}{4}V(0,(c_{2}-c_{1})\sin\psi \cos\psi 
\sin\theta,c_{3}\cos\theta)\\
\gamma_{rr} & = & \frac{c_{0}}{V}, \gamma_{\psi \psi}= \frac{1}{16}(c_{3} 
\sin^{2}\theta(c_{1}cos^{2}(\psi)+c_{2} \sin^{2}\psi)), \non \\
\gamma_{\theta \theta} & = & \frac{1}{16}[c_{1}c_{2} \sin^{2}\theta +c_{3} 
\cos^{2}\theta (c_{1} \sin^{2}\psi +c_{2} \cos^{2}\psi)],\non \\
\gamma_{\theta \psi} & = & -\frac{1}{16}[c_{3}(c_{2}-c_{1}) \sin\psi \cos\psi \sin\theta \cos\theta]
\label{game}
\feqr

\noi where $c_{0}=1/\sqrt{r^2+\alpha^2}$, $c_{1}=c_2=\sqrt{r^2+\alpha^2}$, 
and $c_{3}=r^2/\sqrt{r^2+\alpha^2}$. The explicit expressions for $V$, 
$\vec{A}$ and the components of the $\gamma$ metric are:

\beqr
V & = & \frac{4\sqrt{r^2+\alpha^2}}{r^2+\alpha^2 \sin^{2}\theta}\\
\vec{A} & = & \left(0,0,\frac{r^2 \cos\theta}{r^2+\alpha^2 \sin^{2}\theta}
\right)
\label{omca}
\feqr

\beq
\gamma_{rr}=\frac{r^2+\alpha^2 \sin^{2}\theta}{16(r^2+\alpha^2)}, 
\gamma_{\psi \psi}=\frac{r^2}{16} \sin^{2}\theta, \gamma_{\theta \theta}
=\frac{1}{16}(r^2+\alpha^2 \sin^{2}\theta), \gamma_{\theta \psi}=0
\feq{gamcal} 

The space is flat because the Ricci tensor vanishes:

\beq
R_{i\bar{j}}=g^{\bar{k}l}R_{\bar{k}il\bar{j}}=R^{l}_{il\bar{j}}
=-\frac{\partial^2}{\partial Z^i \partial \bar{Z}^j}\ln (\det(g_{i\bar{j}}))=0.
\feq{ric}

\noi since $\det (g_{i\bar{j}})=4$. A manifold equipped with a Ricci flat 
metric, by Yau's proof of the Calabi conjecture, implies that the first 
Chern class must vanish: $c^{R}_{1}(K)=0$ \footnote{The first Chern class
is represented by the differential form $h=R_{i\bar{j}}dZ^i dZ\bar{j}$ and is 
exact, {\it i.e. \/} that $h=d\beta$ for some $\beta$.}. As a result the 
dimensions of the Dolbeaut cohomology groups $H^{(2,0)}(M)$ and $H^{(0,2)}(M)$
are equal to one. 

\section{Manifold structure and topological invariants} 

The global topology of the EH manifold is \cite{eguchi}:

\begin{enumerate}

\item close to $r=c$, the manifold is homotopic to $S^2$ ($M^{EH}\approx R^2
\times S^2$) and has the same Euler characteristic as $S^2$ {\it i.e. \/} 
$\chi=2$,

\item when $r\rightarrow \infty$ the metric approaches a flat metric and the
constant-r hypersurfaces are distorted three-spheres with opposite points
identified with respect to the origin. The group manifold is then $M_{EH}
\approx R \times RP_3$ where $RP_3$ is the real projective space $RP_3=
SO(3)=S^3/ Z_2$ for which $S^3=SU(2)$ is the double covering.

\end{enumerate}

The Gauss-Bonnet theorem states that it is possible to obtain the Euler 
characteristic of a closed Riemannian manifold of even dimension $dimM=2l$ 
from the volume integral of the $2l$-form $\Omega$: 

\beq
\chi_{M^{2l}}=\int_{M^{2l}}\Omega =\frac{(-1)^{l}}{4^{l}\pi^{l}l!}
\int_{M^{2l}} \epsilon_{a_{1}\cdots a_{2l}}R^{a_{1}a_{2}}\wedge \cdots 
\wedge R^{a_{2l-1}a_{2l}}.
\feq{chern}

\noi In four dimensions it takes the familiar form:

\beq
\chi_{M^{4}}=\frac{1}{32\pi^{2}}\int_{M^4}\epsilon_{abcd}R^{ab}\wedge R^{cd}
\feq{euler4}    
 
\noi where the curvature two-form $R^{a}_{b}$ is defined by the spin 
connection one-forms $\omega^{a}_{b}$ as:

\beq
R^{a}_{b}=d\omega^{a}_{b}+\omega^{a}_{c}\wedge \omega^{c}_{b}.
\feq{connect}

It has also been proved that $\Omega$ can be expresed as the exterior 
derivative of a $(2l-1)$-form in $M^{4l-1}$ constructed by the unit tangent 
vectors of $M^{2l}$:

\beq
\Omega=-dD.
\feq{exteri}  

\noi In this way the original integral of $\Omega$ over $M^{2l}$ can be 
performed over a submanifold $U^{2l}$ obtained as the image in $M^{4l-1}$ of 
a continuous unit tangent vector field over $M^{2l}$ with some isolated 
singular points. Applying Stoke's theorem one thus get:

\beq
\chi_{M^{2l}}=\int_{M^{2l}}\Omega=\int_{U^{2l}}\Omega=\int_{\partial U^{2l}}D.
\feq{stoke}

\noi For manifolds with a boundary, the above formula can be generalized to 
include boundary corrections:

\beq
\chi_{M^{2l}}=\int_{M^{2l}}\Omega - \int_{\partial M^{2l}}D=
\int_{\partial U^{2l}} D - \int_{\partial M^{2l}}D.
\feq{bounda}

\noi The 3-form D in four dimensions is given by \cite{eguchi, gibbons}:

\beq
D=-\frac{1}{16\pi^{2}}\epsilon_{abcd} (\theta^{ab}\wedge R^{cd} 
- \frac{2}{3} \theta^{ab}\wedge \theta^{c}_{e}\wedge \theta^{ed})
\feq{3form} 

\noi where $\theta^{a}_{b}$ is the second fundamental form of the Lorentz 
group {\it i.e.\/} the difference between the spin connection of the original 
metric, computed on the boundary, and the spin connection obtained from the 
boundary if the metric were locally a product near the boundary:

\beq
\theta^{a}_{b}=\omega^{a}_{b}-(\omega_{0})^{a}_{b}.
\feq{funda}

For the EH-manifold the boundary $\partial M_{EH}$ will be represented 
by a slice at $r=r_{0}$ which in the end we will have to send to infinity. 
Since the EH metric factorizes on $\partial M_{EH}$ one easily calculates
the components of the second fundamental form to be:

\beqr
\theta_{12}\!\!&=&\!\! \theta_{23}=\theta_{31}=0 \non \\
\theta_{01} & = & -\left(1-(\frac{\alpha}{r_{0}})^{4} \right)^{\frac{1}{2}}
\sigma_{x} \non\\
\theta_{02} & = & -\left(1-(\frac{\alpha}{r_{0}})^{4}
\right)^{\frac{1}{2}}\sigma_{y}\\
\theta_{03} & = & -\left(1+(\frac{\alpha}{r_{0}})^{4}\right) \sigma_{z} \non
\label{ehfunda}
\feqr

\noi From (\ref{bounda}), (\ref{3form}) and (32) it follows that:

\beq
\chi_{EH}=\frac{3}{2}+\frac{1}{2}=2.
\feq{ec4}

\noi The Hirzebruch signature $\tau$ receives contribution only from the 
``bulk'' and is given by: 

\beq
\tau_{EH}=\frac{1}{24\pi^{2}}\int_{M_{EH}} Tr(R\wedge R)
=\frac{1}{12\pi^{2}} \int_{M_{EH}} R_{ab}R_{ba}=1.
\feq{hirz}

\noi An alternative calculation for $\tau_{EH}$ is presented 
in the Appendix B. In general for the multi-Taub NUT metrics the Euler 
characteristic and the Hirzebruch signature are connected through the 
relation $\chi=\tau +1$ where $\tau=n-1$, n being the number of centers.

%%%%%%%%%%%%%%%%%%%%%%%%%%%%%%%%%%%%%%%%%%%%%%%%%%%%%%%%%%%%%%%%%%%%%%%%%%%%%%%%%%%%%%%%%%%
\section{The orbifold limit of the EH-metric} 

In the limit when $\vec{x}_i \rightarrow 0$, or equivalently 
$\vec{x}\rightarrow \infty$ the two-centered metric degenerates to the 
orbifold metric on $M=C^2/\Gamma$ \footnote{The ALE spaces can also be 
described as the complex affine variety \cite{slodowy} in $C^3$ 
characterized by the vanishing locus $W(x,y,z)=0$, where $x, y, z$ are the 
coordinates on $C^3$. For the $A$-series the corresponding discrete 
subgroups of $SU(2)$ are the binary cyclic groups $C_n=Z_{2n}$ whose action 
on $x, y$ is generated by $\left( \begin{array}{c} x\\ y \end{array} 
\right)\rightarrow \left( \begin{array}{cc} e^{i\frac{\pi}{n}} & 0\\ 0 & 
e^{-i\frac{\pi}{n}}\end{array} \right)\left( \begin{array}{c} x\\ y 
\end{array} \right)$. The smallest set which generates all polynomials 
invariant under $C_n$ is clearly $X=x^{2n}$, $Y=y^{2n}$, $Z=xy$ and these 
generators obey the single relation $XY=Z^{2n}$. In the vincinity of the 
Kleinian singularity of $C^2/Z_2$ the real varieties have a double cone 
structure (the tips of the cones face each other).}. $\Gamma$ in general 
will be a Kleinian subgroup of $SU(2)$ (for multi-centered metrics is 
$Z_{k+1}$) but in the Eguchi-Hanson case (two-centered or $k=1$) will be 
identified to be $Z_2$. The limit $\vec{x}_i \rightarrow 0$ is also 
equivalent to taking $a\rightarrow0$ in the line element {\it i.e.\/}

\beqr
ds^{2}_{C^2/Z_{2}} &=& \lim_{a\rightarrow 0}ds^{2}_{E.H.}=\frac{1}{4r}dr^2
+r(\sigma^{2}_{x}+\sigma^{2}_{y}+\sigma^{2}_{z}) \non\\
&=& \frac{1}{4r}dr^2 + \frac{r}{4}(d\phi^2 + d\theta^2 + d\psi^2 
+2\cos^{2}\theta d\psi d\phi ).
\label{orbif}
\feqr

\noi When we approach the origin we encounter a conical singularity which 
corresponds to a point of higher symmetry \footnote{In this case both vector 
fields $\frac{\partial}{\partial\psi}$ and $\frac{\partial}{\partial\phi}$ 
generate translational isometries and since flat space is self-dual as well 
as anti-self-dual the rotational character of $\frac{\partial}{\partial\psi}$ 
or $\frac{\partial}{\partial\phi}$ disappears.}. The self-dual components of 
the spin connection are easily computed to be: 

\beqr
\omega^{1}_{0}=\omega^{2}_{3}=\sigma_{x}=\frac{e^1}{r} \non \\
\omega^{2}_{0}=\omega^{3}_{1}=\sigma_{y}=\frac{e^2}{r} \\
\omega^{3}_{0}=\omega^{1}_{2}=\sigma_{z}=\frac{e^3}{r} \non
\label{orsp}
\feqr

\noi where the orthonormal vierbein basis is $e^{a}=(e^0,e^1,e^2,e^3)
=(dr,r\sigma_{x},r\sigma_{y},r\sigma_{z})$. A factor of 4r has been absorbed 
in the line element {\it i.e. \/} 

\beq
d\tilde{s}^2=dr^2+r^2(\sigma^{2}_{x}+\sigma^{2}_{y}+\sigma^{2}_{z}).
\feq{orab}

\noi From the definition of the curvature two-form we find that its 
components vanish so this metric is flat everywhere apart from the fixed 
point.

In terms of our original expression (eq. (\ref{kahm})) for the K\"ahler metric 
the orbifold limit corresponds to taking both $c,r$ to zero. We can express
$V$, $\vec{A}$ and $\gamma_{ij}$ of the line 
element in terms of the $Z^1$, $Z^2$ coordinates as follows (see Appendix B): 

\beqr
V_{orb}=\frac{4}{|Z^{1}|^{2}+|Z^{2}|^{2}}, \,\,\,\, A_{orb}
=\pm \frac{|Z^{1}|^{2}-|Z^{2}|^{2}}{|Z^{1}|^{2}+|Z^{2}|^{2}} \non\\
\gamma_{rr}=\frac{1}{16}, \,\,\, \gamma_{\psi \psi}
= \frac{1}{4}|Z^{1}|^{2}|Z^{2}|^{2}, \,\,\, \gamma_{\theta \theta}
=\frac{1}{16}(|Z^{1}|^{2}+|Z^{2}|^{2})^{2}, \,\,\, \gamma_{\theta \psi}=0.
\label{vorb}
\feqr

The ``bulk'' contributions to the Euler characteristic and the Hirzebruch 
signature vanish since $R_{ab}=0$ so we end up with:

\beqr
\chi_{C^2/Z_{2}}=1\non \\
\tau_{C^2/Z_{2}}=0.
\label{orbe}
\feqr

\noi These results are also confirmed by the prementioned formula of the 
Euler characteristic expressed by $\chi=\tau+1=n$.

%%%%%%%%%%%%%%%%%%%%%%%%%%%%%%%%%%%%%%%%%%%%%%%%%%%%%%%%%%%%%%%%%%%%%%%%%%%%%%%%%%%%%
\section{Classical motion on the $C^2/ Z_{2}$ orbifold}

Consider the classical wave equation of a particle with mass $\mu$ on the 
quotient manifold $M=C^{2}/Z_{2}$ described by:

\beq
\partial_{\mu}(\sqrt{g}g^{\mu\nu}\partial_{\nu}F)-\sqrt{g}\mu^{2}F=0
\feq{wavee}

\noi where $F=F(r,\theta,\phi,\psi)=R(r)Q(\theta, \phi, \psi)$ 
and $g=\det{g_{ij}}=r^{6}\sin^{2}\theta$. The radial equation takes the form: 

\beq
\frac{d^2 R}{dr^{2}}+\frac{(q-1)}{r}\frac{dR}{dr}
-( \mu^{2}+\frac{\lambda^2}{r^2})R=0
\feq{radwa}

\noi with $\lambda^{2}=l(l+1)$ a positive constant determining the opening 
angle and $q=dimM$ is the dimension of the manifold. Changing variable to 
$\rho=r\mu$ we obtain the generalized Bessel's equation \cite{watson}:

\beq
\frac{d^2 J}{d\rho^2}+\frac{(q-1)}{\rho}\frac{dJ}{d\rho}-
(1+\frac{\lambda^2}{\rho^2})J=0.
\feq{besl}

\noi Solving (\ref{besl}) in the standard way we find the solution for the  
$(s^{+} - s^{-})\notin Z^{+}$ case to be (see Appendix C for the 
other solutions):

\beqr
J(\rho) &=& AJ_{s^{+}}(\rho)+BJ_{s^{-}}(\rho) \non \\
&=& A\left(\frac{\rho}{2}\right)^{s^{+}} \sum_{j=0}^{\infty}
\left(\frac{\rho}{2}\right)^{2j}
\frac{1}{j! \Gamma (1+j+d)}+B\left(\frac{\rho}{2}\right)^{s^{-}} 
\sum_{j=0}^{\infty}\left(\frac{\rho}{2}\right)^{2j}\frac{1}
{j! \Gamma (1+j+d)} 
\label{solbel}
\feqr
 
\noi where $s^{\pm}=s^{\pm}_{(l,q)}=-\frac{(q-2)}{2} \pm d
=-\frac{(q-2)}{2} \pm \sqrt{\lambda^2 + (\frac{(q-2)}{2})^2}$, $i=1,2$. 
The limiting forms of the Bessel's
functions for small and large values of their arguments are given by 
the leading terms:

\beqr
\rho \ll 1 &,& J_{s^{\pm}}(\rho)\rightarrow \frac{1}{\Gamma(s^{\pm} 
+ \frac{q}{2})} \left(\frac{\rho}{2}\right)^{s^{\pm}} \\
\rho \gg 1 &,& J_{s^{\pm}}(\rho)\rightarrow \frac{2^{\frac{q-3}{2}}}
{\sqrt{\pi \rho^{q-1}}}\cos \left(\rho-\frac{s^{\pm} \pi}{2} 
- \frac{(q-1)\pi}{4}\right).
\label{asbs}
\feqr

\noi Notice that for $\rho \ll 1$ the solutions are always square integrable
around the singular point $r=0$ and there is a tendency of the wave function 
to concentrate around $r=0$ when $l$ decreases.  

The azimuthal equation has solutions the eigenfunctions of the operators 
$L^2$, $L_{3}$, $\tilde{L}^2$ and $\tilde{L}_3$ {\it i.e.\/}

\beqr
L^2|l,m,n> & = & l(l+1)|l,m,n> \non\\
L_{3}|l,m,n> & = & n|l,m,n> \non\\
\tilde{L}^2|l,m,n> & = & l(l+1)|l,m,n> \non\\
\tilde{L}_{3}|l,m,n> & = & m|l,m,n>
\label{eigenw}  
\feqr

\noi where $|l,m,n>=Q^{l}_{m,n}(\theta,\phi,\psi)
=\Phi_{m}(\phi) \Psi_{n}(\psi) W^{l}_{m,n}(\theta)
=e^{im\phi}e^{in\psi}W^{l}_{m,n}(\theta)$ and $|m|\leq l$, $|n|\leq l$. 
The generators $L_i$ obey the $SU(2)$ algebra $[L_i,L_j]
=i\epsilon_{ij}^{\,\,\,\,k}L_k$, $[L^2,L_3]=0$, while their dual partners the 
commutation relations $[\tilde{L}_i,\tilde{L}_j]
=-i\epsilon_{ij}^{\,\,\,\,k}\tilde{L}_k$, $[\tilde{L}^2,\tilde{L}_{3}]=0$. 
The first equation of (\ref{eigenw}) can be written equivalently as:

\beq
\frac{1}{\sin\theta}\frac{d}{d\theta}(\sin\theta \frac{dW}{d\theta})
+\left[l(l+1)-\frac{m^2 +n^2}{\sin^2 \theta}
+2\frac{\cos\theta}{\sin^2 \theta}mn \right]W=0
\feq{azwon} 

\noi and substituting $u=\cos\theta$ for $\theta$ as well as putting 
$W(\theta)=P(u)$, (\ref{azwon}) becomes: 

\beq
\frac{d}{du}\left[(1-u^2)\frac{dP}{du}\right]+\left[l(l+1)
-\frac{m^2+n^2}{1-u^2}
+2\frac{u}{1-u^2}mn \right]P=0.
\feq{trawon}

\noi The solutions of the differential eq. (\ref{trawon}) have the following 
symmetries:

\begin{enumerate}
\item they are invariant under the interchange of m, n:

\beq
P^{l}_{m,n}(u)=P^{l}_{n,m}(u)=P^{l}_{-m,-n}(u)
\feq{sym1}

\item we recover the familiar result of $\textit{the associated Legendre 
functions of the first kind}$ when $m=0$ or $n=0$:

\beq
P^{l}_{m,0}(u)=P^{l}_{m}(u)=P^{l}_{0,n}(u)=P^{l}_{n}(u).
\feq{sym2}
\end{enumerate}

The singular regular points of eq. (\ref{trawon}) are $u=\mp 1$. Near the 
points $u=\mp 1$ the dominant behaviour is $(\frac{u-1}{2})^{a/2}$ and 
$(\frac{1+u}{2})^{b/2}$ respectively, where $a=|m-n|$ and $b=|m+n|$. Consider 
now the expansion of the form:

\beq
P^{l}_{m,n}(u)=N^{l}_{m,n} \left(\frac{u-1}{2}\right)^{a/2} 
\left(\frac{1+u}{2}\right)^{b/2} 
U^{l}_{m,n}(u)
\feq{exjac1}

\noi with $N^{l}_{m,n}$ a normalization factor to be determined later on and 
$U^{l}_{m,n}(u)$ satisfying the differential equation:

\beq
(1-u^2)\frac{d^2 U}{du^2}+\left[b-a-(2+a+b)\right]\frac{dU}{du}
+\left[l(l+1)-\frac{a+b}{2}(\frac{a+b}{2}-1)\right]U.
\feq{exjac2}

\noi We recognize (\ref{exjac2}) to be a hypergeometric equation with solutions 
$\textit{the Jacobi polynomials}$ \cite{szego}, provided that $l=\tilde{\rho} 
+ \frac{a+b}{2}$ and $\tilde{\rho}\in Z$. One can prove that equation:

\beq
\frac{d}{du}\left[(1-u)^{a+1}(1+u)^{b+1} \frac{dU}{du}\right]
+\tilde{\rho} (\tilde{\rho} +a+b+1)U=0
\feq{exjac3}

\noi has solutions given by $\textit{the Rodrigue's formula}$: 

\beqr
T^{(a,b)}_{k}(u) &=& \frac{(-1)^k}{2^k k!}(1-u)^{-a}(1+u)^{-b} 
\frac{d^k}{d u^k}\left[(1-u)^{k+a}(1+u)^{k+b}\right] \non \\
&=& \sum_{\nu=0}^{k} \left( \begin{array}{c} k+a\\ k-\nu \end{array} 
\right) \left( \begin{array}{c} k+b\\ \nu \end{array} 
\right)\left(\frac{u-1}{2}\right)^{\nu} \left(\frac{1+u}{2}\right)^{k-\nu}
\label{jacobi}
\feqr

\noi and obeying the important identity:

\beq
T^{(a,b)}_{k}(u)=(-1)^k T^{(b,a)}_{k}(-u).
\feq{jacoid}

\noi In our case the solutions \footnote{
It was shown in \cite{golberg} that 
in complex stereographic coordinates $(\zeta,\bar{\zeta})$, defined by 
$\zeta=e^{i\phi}\cot \frac{\theta}{2}$, \\ $\bar{\zeta}=e^{-i\phi}\cot 
\frac{\theta}{2}$ the spin-s spherical harmonics take 
the form $Y^{l}_{m,s}=\left[ 
\frac{(2l+1)(l-m)!(l+m)!}{4\pi(l-s)!(l+s)!} \right]^{\frac{1}{2}}
(-1)^{l-m}$  \\ $\times (1+\zeta\bar{\zeta})^{-1}  
\sum_{\nu} \left( \begin{array}{c} l-s\\ p \end{array}\right)
\left( \begin{array}{c} l+s\\ p+s-m \end{array} \right) \zeta^p 
(-\bar{\zeta})^{p+s-m}$. In Appendix D we prove that these solutions are 
related to the Jacobi polynomials up to a symmetry.} are given explicitly 
by:

\beqr
P^{l}_{m,n}(u) &=& e^{in\psi}Y^{l}_{m,n}(\theta,\phi)\non\\
&=& \left[\frac{2l+1}{4\pi}\frac{(l-m)!(l+m)!}{(l-n)!(l+n)!} 
\right]^{\frac{1}{2}} e^{im\phi} e^{in\psi} T^{(a,b)}_{l-m} 
\left( \frac{1-u}{2}\right)^{\frac{a}{2}} \left(\frac{1+u}{2} 
\right)^{\frac{b}{2}} 
\label{soljac}
\feqr

\noi and form a complete set of orthonormal polynomials {\it i.e.\/}:

\beqr
\int_{0}^{2\pi}d\phi\int_{0}^{2\pi}d\psi \int_{-1}^{+1} du \,\, 
Q^{l\ast}_{m,n}(u,\phi,\psi) Q^{l^{\prime}}_{m^{\prime},n^{\prime}}
(u,\phi,\psi)= \frac{8\pi^2}{2l+1}\delta_{l,l^{\prime}} 
\delta_{m,m^{\prime}} \delta_{n,n^{\prime}} \\
\sum_{l,m,n}Q^{l\ast}_{m,n}(\theta,\phi,\psi)Q^{l}_{m,n}
(\theta^{\prime},\phi^{\prime},\psi^{\prime})=\frac{8\pi^2}{2l+1}
\delta(\cos \theta-\cos\theta^{\prime})\delta(\phi-\phi^{\prime})
\delta(\psi-\psi^{\prime}).
\label{normjac}
\feqr

%%%%%%%%%%%%%%%%%%%%%%%%%%%%%%%%%%%%%%%%%%%%%%%%%%%%%%%%%%%%%%%%%%%%%%%%%%%%%%%%%%%%%%%%%%%%%%%%%%

\section{Schr\"{o}dinger's equation on $C^2/Z_2$.}

In the presence of the harmonic potential $V(r)=\frac{1}{2}Kr^2$ 
Schr\"{o}dinger's equation becomes:

\beq
\frac{\hbar^2}{2\mu}\left[\frac{d^2 R}{dr^2}+\frac{(q-1)}{r}\frac{dR}{dr} 
\right]-\left[2(V-E)+\frac{2\mu}{\hbar^2}\frac{l(l+1)}{r^2}\right]R=0
\feq{scheq}

\noi which by changing variable to $\xi=\left(\frac{2\mu K}{\hbar^2}\right)
^{\frac{1}{4}}r$ and introducing the  quantities $a^4=\frac{2\mu K}{\hbar^2}$ and 
$\lambda=\frac{2E}{\hbar}\left(\frac{2\mu}{K}\right)^{\frac{1}{2}}$ it can 
be rewritten in the following dimensionless form as:

\beq
\frac{d^2 R}{d\xi^2}+\frac{(q-1)}{\xi}\frac{dR}{d\xi}+\left[\lambda
-\xi^2-\frac{l(l+1)}{\xi^2}\right]R=0.
\feq{harmsch}

\noi The solution of eq. (\ref{harmsch}) is facilitated on one hand by examining 
the dominant behaviour of $R(\xi)$ in the asymptotic region 
$\xi\rightarrow +\infty$ and on the other hand by demanding finiteness at 
$\xi=0$. Then the desired solution is: 

\beq
R(\xi)=e^{-\frac{1}{2}\xi^2} \xi^{s^{+}} L(\xi)
\feq{sfin}

\noi where $L(\xi)$ satisfies the equation:

\beq
\xi L^{\prime \prime}+(2s^{+}+q-1-2\xi^2)L^{\prime}+2\nu\xi L=0.
\feq{Lequ}

\noi Energy levels are determined in the usual way and are found to be: 

\beq
E_{l,\nu,q}=\frac{\hbar\omega_{c}}{\sqrt{2}}\left[\nu+1+\sqrt{\frac{(q-2)^2}
{4}+l(l+1)} \right]
\feq{enespe}

\noi with $\nu$ being an integer ($\nu\geq 0$) and $\omega_{c}
=\sqrt{\frac{K}{\mu}}$ the angular frequency of the corresponding classical 
harmonic oscillator. In one dimension and when $l=0$ we obtain the expected 
result.

The solutions (\ref{sfin}) fall into two classes depending on whether $\nu$ is 
even or odd integer.

\begin{enumerate}
\item $\nu=2n$: 

\beq
R_{even}(\xi)=e^{-\frac{1}{2} \xi^2} \xi^{s^{+}} \left[ \frac{\Gamma(-n)}
{\Gamma(d)} + \sum_{k=1}^{2(n-1)} \xi^{2k} \frac{\Gamma(k-n)}
{\Gamma(k) \Gamma(1+k+d)}\right]
\feq{sol1}

\item $\nu=2n+1$: 

\beq
R_{odd}(\xi)=e^{-\frac{1}{2} \xi^2} \xi^{s^{+}} \left[ 
\frac{\Gamma(-\frac{1}{2}-n)}
{\Gamma(d)} + \sum_{k=1}^{2n-1} \xi^{2k} \frac{\Gamma(k-n-\frac{1}{2})}
{\Gamma(k) \Gamma(1+k+d)}\right],
\feq{sol2}

\end{enumerate}

\noi where $\Gamma(-z)=-\frac{\pi}{z\sin \pi z \Gamma(z)}$. 

For the Coulomb potential $V(r)=-\frac{a}{r}$ inserting the quantity
$\lambda=\frac{a}{\hbar}\left( \frac{\mu}{E^{\prime}} \right)^{\frac{1}{2}}$
(where $E^{\prime}=|E|$) and the new variable
$\xi=\frac{4a\mu}{\hbar^2\left[\nu+\frac{1}{2}
+\sqrt{\frac{(q-2)^2}{4}+l(l+1)} \right]} r$ into Schr\"{o}dinger's equation 
we get:

\beq
\frac{d^2 R}{d\xi^2}+\frac{(q-1)}{\xi}\frac{dR}{d\xi}
+\left[\frac{\lambda}{\xi}-\frac{1}{4}-\frac{l(l+1)}{\xi^2}\right]R=0.
\feq{Coulpo} 

\noi Again the energy spectrum is specified by the asymptotic behaviour of 
the power series $L(\xi)$ and is:

\beq
E_{l,\nu,q}=-\frac{a^2 \mu}{\hbar^2}\frac{1}{\left[\nu+\frac{1}{2}
+\sqrt{\frac{(q-2)^2}{4}+l(l+1)} \right]^2}
\feq{enecoul}

\noi The solutions satisfying the same boundary conditions as in the case of 
the harmonic potential are:

\beq
R_{\nu,l,q}(\xi)=e^{-\frac{1}{2}\xi}\xi^{s^{+}} L^{2s^{+}+q-2}_{\nu}
(\xi)
\feq{laguer} 

\noi where $L^{2s^{+}+q-2}_{\nu}(\xi)$ are 
$\textit{Laguerre's polynomials}$ defined by the series:

\beq
L^{2s^{+}+q-2}_{\nu}(\xi)=\sum_{k=0}^{\nu}\frac{\Gamma(\nu+2s^{+}
+q-1)}{\Gamma(k+2s^{+}+q-1)}\frac{(-\xi)^k}{k!(\nu-k)!}
\feq{Lagupol}

\noi with $2s^{+}+q-2=\nu+\frac{1}{2}+\sqrt{\frac{(q-2)^2}{4}+l(l+1)}\in 
R$. We can easily check from (\ref{Lagupol}) that in three dimensions we 
obtain the correct results.

%%%%%%%%%%%%%%%%%%%%%%%%%%%%%%%%%%%%%%%%%%%%%%%%%%%%%%%%%%%%%%%%%%%%%%%%%%

\section{Conclusions.} 

We have presented a simple way to derive the Eguchi-Hanson metric from
the $N=2$ action of nonlinear $\sigma$-models. The Hirzebruch signature  
of the manifold has been calculated by using the K\"{a}hler structure 
that accompanies it. A simple way has been proposed to construct the orbifold
$M=C^2 / Z_{2}$ from the EH metric as well as the study of its topological
invariant features have been explored. The classical and the quantum 
behaviour of a massive particle on the singular manifold has been investigated on
one hand by studying the classical wave equation and on the other hand 
by Schr\"{o}dinger's equation and switching on some potentials.

%%%%%%%%%%%%%%%%%%%%%%%%%%%%%%%%%%%%%%%%%%%%%%%%%%%%%%%%%%%%%%%%%%%%%%%%%%

\section{Acknowledgements}

I would like to thank M. Ro\v{c}ek for enlightening discussions from which
I benefited greatly.

%%%%%%%%%%%%%%%%%%%%%%%%%%%%%%%%%%%%%%%%%%%%%%%%%%%%%%%%%%%%%%%%%%%%%%%%%

\section{Appendix A}

\renewcommand{\theequation}{A.\arabic{equation}}
\setcounter{equation}{0}

The left-invariant one forms $\sigma_i$ on the manifold of the group 
$SU(2)_L=S^3$, have the explicit expression:

\beqr
\sigma_{x} & = & \frac{1}{2}(\sin \psi d \theta - \cos\psi \sin\theta 
d \phi)\non\\
\sigma_{y} & = & \frac{1}{2}(\sin\psi\sin\theta d\phi-\cos\psi 
d\theta)\non\\
\sigma_{z} & = & \frac{1}{2}(d\psi+\cos\theta d\phi).
\label{dcatil}
\feqr

\noi and obey the relation:

\beq
d\sigma_i=\epsilon_{ijk} \sigma_{j} \wedge \sigma_{k}.
\feq{difsig}
  
\noi In an analogous way one can define the dual one-forms $\tilde{\sigma}_i$ 
invariant under the $SU(2)_{R}$ supersymmetry and in terms of Euler angles 
they read:

\beqr
\tilde{\sigma}_{x} & = & \frac{1}{2}(\sin \phi d\theta - \cos\phi \sin\theta 
d\psi)\non\\
\tilde{\sigma}_{y} & = & \frac{1}{2}(\sin\phi\sin\theta d\psi-\cos\phi 
d\theta)\non\\
\tilde{\sigma}_{z} & = & \frac{1}{2}(d\phi+\cos\theta d\psi).
\label{dcartm}
\feqr

\noi The generators of $SU(2)$ are expressed in terms of the killing vectors 
$\xi^{\mu}_{(j)}$ of the metric by the formula $L_j
=-i\xi^{\mu}_{(j)}\partial_{\mu}$. In spherical coordinates they read:

\beqr
L_1 & = & -i\left(\sin \psi \frac{\partial}{\partial \theta}-\frac{\cos \psi}
{\sin \theta}\frac{\partial}{\partial \phi}+\frac{\cos \theta \cos \psi}
{\sin \theta}\frac{\partial}{\partial \psi} \right) \non\\
L_2 & = & i\left(\cos \psi \frac{\partial}{\partial \theta}+\frac{\sin \psi}
{\sin \theta}\frac{\partial}{\partial \phi}-\frac{\cos \theta \sin \psi}
{\sin \theta}\frac{\partial}{\partial \psi} \right) \non\\
L_3 & = & -i\frac{\partial}{\partial\psi}
\label{opl}
\feqr

\noi where $L_3$ is associated with a diagonal $U(1)$. Similar expressions 
hold for the dual generators provided we interchange the roles of 
$\phi$ and $\psi$.

%%%%%%%%%%%%%%%%%%%%%%%%%%%%%%%%%%%%%%%%%%%%%%%%%%%%%%%%%%%%%%%%%%%%%%%%%%%%%%%

\section{Appendix B}

\renewcommand{\theequation}{B.\arabic{equation}}
\setcounter{equation}{0}
  
Let $M$ be a complex manifold with local coordinates $(Z^{1},Z^{2},\cdots ,
Z^{n})$. A Hermitian metric on $M$ is given by an expression of the form:

\beq
g_{a\bar{b}}dZ^a\otimes d\bar{Z}^b.
\feq{clo}

\noi That $g_{a\bar{b}}$ is Hermitian means: 

\beq
g_{a\bar{b}}=\bar{g}_{b\bar{a}}.
\feq{her}

\noi We also require:

\beq
g_{a\bar{b}}=g_{\bar{b}a}, g_{ab}=0=g_{\bar{a}\bar{b}}.
\feq{herz}

\noi The affine connection and the Riemann tensor greatly simplify on a 
K\"{a}hler manifold as one can realise. The K\"{a}hler 2-form $K
= g_{a\bar{b}}dz^a \wedge d\bar{z}^b$ is closed ($dK=0$) and this implies 
that:

\beq
\partial_{c} g_{a\bar{b}}=\partial_{a} g_{c\bar{b}}, \partial_{\bar{c}} 
g_{a\bar{b}}=\partial_{\bar{b}} g_{a\bar{c}}. 
\feq{equl}

\noi The affine connection is given as usual by $\Gamma_{ij,k}
=(\partial_{i}g_{jk}+\partial_{j}g_{ik}-\partial_{k}g_{ij})/2$ and taking 
into account (\ref{clo}) one finds that the only nonzero components are:

\beq
\Gamma_{ab,{\bar{c}}}=\partial_{b}g_{a\bar{c}}, \Gamma_{\bar{a}\bar{b},c}
=\partial_{\bar{b}}g_{\bar{a}c}.
\feq{chri}

\noi For the covariant Riemann curvature the definition:

\beq
R_{ijkl}=\frac{1}{2}(\frac{\partial^{2} g_{ik}}{\partial x^{j}
\partial x^{l}}-\frac{\partial^{2} g_{il}}{\partial x^{j}\partial x^{k}}
-\frac{\partial^{2} g_{jk}}{\partial x^{i}\partial x^{l}}+\frac{\partial^{2} 
g_{jl}}{\partial x^{i}\partial x^{k}})+g_{\eta\rho}(\Gamma^{\eta}_{ik}
\Gamma^{\rho}_{jl}-\Gamma^{\eta}_{il}\Gamma^{\rho}_{jk})
\feq{riem}

\noi becomes:

\beq
R_{a\bar{b}c\bar{d}}=-\frac{1}{2}(\partial_{\bar{b}}\Gamma_{ac,\bar{d}}
+\partial_{\bar{d}}\Gamma_{ac,\bar{b}})-g^{\bar{\tau}\rho}
\Gamma_{ac,\bar{\tau}}\Gamma_{\bar{b}\bar{d},\rho}.
\feq{riex}

\noi The above expression provides the nonzero components which turn out to 
be:

\beq
R_{a\bar{b}c\bar{d}}, R_{\bar{a}b\bar{c}d}, R_{a\bar{b}\bar{c}d}, 
R_{\bar{a}bc\bar{d}}.
\feq{comp} 

\noi $\textit{The first Bianchi identity}$:

\beq
R_{ijkl}+R_{kijl}+R_{jkil}=0
\feq{fbian}

\noi simplifies to the statement that:

\beq
R_{a\bar{b}c\bar{d}}=R_{c\bar{d}a\bar{b}}
\feq{nbian}

\noi while $\textit{the second Bianchi identity}$ becomes:

\beq
R_{a\bar{b}c\bar{d},f}+R_{\bar{b}fc\bar{d},a}+R_{fac\bar{d},\bar{b}}=0.
\feq{sbian}

\noi The nonzero components of the affine connnection for the metric 
(\ref{kahm}) are: 

\beqr
\Gamma_{\bar{1}\bar{1},1} & = & \partial_{\bar{1}}g_{1\bar{1}}
=\frac{(2\mid Z_{1}\mid^6-6\mid Z_{1}\mid^2 \mid Z_{2}\mid^4 
-\mid Z_{2}\mid^2 c^2 -4\mid Z_{2}\mid^6 )Z_{1}c^2}
{4r^6(1+\frac{c^2}{4r^2})^{3/2}} \\
\Gamma_{\bar{1}\bar{2},1} & = & \partial_{\bar{2}}g_{1\bar{1}} \non\\ 
&=& \frac{(8\mid Z_{1}\mid^6+12\mid Z_{1}\mid^4 \mid Z_{2}\mid^2 
+\mid Z_{1}\mid^2 c^2 -4\mid Z_{2}\mid^6 
-\mid Z_{2}\mid^2 c^2)Z_{2}c^2}{8r^6(1+\frac{c^2}{4r^2})^{3/2}} \\
\Gamma_{\bar{2}\bar{2},2} & = & \partial_{\bar{2}}g_{2\bar{2}}
=\frac{(2\mid Z_{2}\mid^6-6\mid z_{1}\mid^4 \mid Z_{2}\mid^2 
-\mid Z_{1}\mid^2 c^2 -4\mid Z_{1}\mid^6 )Z_{2}c^2}
{4r^6(1+\frac{c^2}{4r^2})^{3/2}} \\
\Gamma_{\bar{2}\bar{1},2} & = & \partial_{\bar{1}}g_{2\bar{2}} \non\\
&=& \frac{(8\mid Z_{2}\mid^6+12\mid Z_{1}\mid^2 \mid Z_{2}\mid^4 
+\mid Z_{2}\mid^2 c^2 -4\mid Z_{1}\mid^6 -\mid Z_{1}\mid^2 c^2)
Z_{1}c^2}{8r^6(1+\frac{c^2}{4r^2})^{3/2}} \\
\Gamma_{\bar{1}\bar{1},2} & = & \partial_{\bar{1}}g_{2\bar{1}}
=\frac{(6\mid Z_{1}\mid^4+12\mid Z_{1}\mid^2 \mid Z_{2}\mid^2 
+6\mid Z_{2}\mid^4+ c^2)Z^{2}_{1}\bar{Z}_{2} c^2}
{4r^6(1+\frac{c^2}{4r^2})^{3/2}} \\
\Gamma_{\bar{2}\bar{2},1} & = & \partial_{\bar{2}}g_{1\bar{2}}
=\frac{(6\mid Z_{1}\mid^4+12\mid Z_{1}\mid^2 
\mid Z_{2}\mid^2 +6\mid Z_{2}\mid^4+c^2)\bar{Z}_{1} 
Z^{2}_{2}c^2}{4r^6(1+\frac{c^2}{4r^2})^{3/2}}.
\label{chrn}
\feqr

The Weyl tensor in our case equals the Riemann curvature and the invariant 
we consider is the quadratic expression:

\beq
W=R^{\bar{i}j\bar{k}l} R_{\bar{i}j\bar{k}l}.
\feq{inv}

\noi The nonzero components of the Weyl tensor are (excluding components 
connected through obvious symmetries):

\beq
R_{1\bar{1}1\bar{1}}, R_{2\bar{2}2\bar{2}}, R_{1\bar{1}1\bar{2}}, 
R_{1\bar{1}2\bar{1}}, R_{2\bar{2}2\bar{1}}, R_{2\bar{2}1\bar{2}}, 
R_{1\bar{1}2\bar{2}}, R_{1\bar{2}1\bar{2}}, R_{1\bar{2}2\bar{1}}, 
R_{2\bar{1}2\bar{1}}
\feq{wey}

\noi and the quadratic invariant thus takes the form:

\beqr
W &=& 4R^{\bar{1}1\bar{1}1} R_{\bar{1}1\bar{1}1}+4R^{\bar{2}2\bar{2}2} 
R_{\bar{2}2\bar{2}2}+8R^{\bar{1}1\bar{1}2} R_{\bar{1}1\bar{1}2} \non \\
&+& 8R^{\bar{1}1\bar{2}1} R_{\bar{1}1\bar{2}1}+8R^{\bar{2}2\bar{2}1} 
R_{\bar{2}2\bar{2}1} +8R^{\bar{2}2\bar{1}2} R_{\bar{2}2\bar{1}2} \non \\
&+& 8R^{\bar{1}1\bar{2}2} R_{\bar{1}1\bar{2}2} +4R^{\bar{1}2\bar{1}2} 
R_{\bar{1}2\bar{1}2}+8R^{\bar{1}2\bar{2}1} R_{\bar{1}2\bar{2}1}+
4R^{\bar{2}1\bar{2}1} R_{\bar{2}1\bar{2}1}.
\label{weyin}
\feqr

\noi For the metric (\ref{kahm}) the result reads:

\beq
W=\frac{384c^4}{(4r^2+c^2)^3}
\feq{inka}

\noi which vanishes when $c=0$. Integration over the whole $SO(3)$ sphere 
gives:

\beq
\frac{1}{24\pi^2}\int Tr(R \wedge R)= 1.
\feq{hierz}

\noi The inverse transformations read:

\beqr
r=\mid Z^{1}\mid^{2}+\mid Z^{2}\mid^{2}&,&\,\cos \frac{\theta}{2}
=\frac{\mid Z^{1}\mid}{\sqrt{\mid Z^{1}\mid^{2}+\mid Z^{2}\mid^{2}}},\,\, 
\sin\frac{\theta}{2}=\frac{\mid Z^{2}\mid}{\sqrt{\mid Z^{1}\mid^{2}
+\mid Z^{2}\mid^{2}}}, \non \\
\psi&=&-i\ln \left(\frac{Z^{1}Z^{2}}{\mid Z^{1}\mid \mid Z^{2}\mid}
\right),\,\, 
\phi=i\ln \left(\frac{Z^{2}\mid Z^{1} \mid}{Z^{1} \mid Z^{2} \mid}\right)
\label{invtr}
\feqr

\noi bearing in mind that $0\leq \theta \leq \pi$, $0\leq \phi \leq 2\pi$, 
$0\leq \psi \leq 2\pi$.

%%%%%%%%%%%%%%%%%%%%%%%%%%%%%%%%%%%%%%%%%%%%%%%%%%%%%%%%%%%%%%%%%%%%%%%%%%%%%%%%
\section{Appendix C}

\renewcommand{\theequation}{C.\arabic{equation}}
\setcounter{equation}{0}

The other two solutions of Bessel's equation are:

\begin{enumerate}

\item $(s^{+} - s^{-})\in Z^{+}$:

\beq
J(\rho)=AJ_{s^{+}} (\rho)+B[J_{s^{-}}(\rho)+CJ_{s^{+}}(\rho)\ln 
\mid \rho \mid]
\feq{slbes2}

\item $s^{+} - s^{-}=0$:

\beq
J(\rho)=AJ_{s^{+}}(\rho)+B[\rho^{s^{+} +1}\sum_{j=0}^{\infty} 
\alpha^{\ast}_{2}\rho^{2j}+J_{s^{+}}(\rho)\ln \mid \rho \mid].
\feq{solbel3}

\end{enumerate}

\noi The roots of the Bessel's function are given by the asymptotic formula 
of $J_{s^{\pm}}(\rho)$:

\beq
\rho^{\pm}_{\lambda, q,n}=n\frac{\pi}{2}+\frac{\pi}{2}
\left[\frac{1}{2} \pm \sqrt{\lambda^2 + (\frac{q-2}{2})^2}\right].
\feq{roots}

\noi The normalization of Bessel's function is:

\beq
\int_{0}^{\alpha}\rho J_{s^{+}}(x_{s^{+}},q,n \frac{\rho}{\alpha})
J_{s^{+}}(x_{s^{+}},q,n^{\prime} \frac{\rho}{\alpha}) d\rho=
\frac{\alpha^2}{2}J^{2}_{s^{+} +1}(x_{s^{+}},q,n \delta_{n,n^{\prime}}).
\feq{norbe}

%%%%%%%%%%%%%%%%%%%%%%%%%%%%%%%%%%%%%%%%%%%%%%%%%%%%%%%%%%%%%%%%%%%%%%%%%%%%%%%%%%

\section{Appendix D}

\renewcommand{\theequation}{D.\arabic{equation}}
\setcounter{equation}{0}

The $\textit{generating function}$ of the Jacobi polynomials is: 

\beqr
\sum_{k=0}^{\infty}T^{(a,b)}_{k}(u)w^k &=& 
2^{a+b}(1-2uw+w^2)^{-\frac{1}{2}}[1-w+(1-2uw+w^2)^{\frac{1}{2}}]^{-a} \non\\
&\times& [1+w+(1-2uw+w^2)^{\frac{1}{2}}]^{-b}
\label{genj}
\feqr

\noi where the expressions $[\cdots]^{-a}$ and $[\cdots]^{-b}$ must be taken 
positive for $w=0$.

The recurrence relations are:

\beqr
uT^{(a,b)}_{k}(u) &=& \frac{2(k+a)(k+b)}{(2k+a+b)(2k+a+b+1)}
T^{(a,b)}_{k-1}(u) \non\\
&-& \frac{(a^2 -b^2)}{(2k+a+b)(2k+a+b+2)}T^{(a,b)}_{k}(u)\non\\
&+& \frac{2(k+1)(k+a+b+1)}{(2k+a+b+1)(2k+a+b+2)}T^{(a,b)}_{k+1}(u)
\label{recuj1}
\feqr

\noi and:

\beqr
(1-u^2)\frac{d T^{(a,b)}_{k}(u)}{d u} &=& \frac{2(k+a)(k+b)(k+a+b+1)}
{(2k+a+b)(2k+a+b+1)}T^{(a,b)}_{k-1}(u) \non\\
&+& \frac{2k(a-b)(k+a+b+1}{(2k+a+b)(2k+a+b+2)}T^{(a,b)}_{k}(u) \non\\
&-& \frac{2k(k+1)(k+a+b+1)}{(2k+a+b+1)(2k+a+b+2)}T^{(a,b)}_{k+1}(u)
\label{recuj2}
\feqr

\noi where $k=2,3,\cdots$, $T^{(a,b)}_{0}(u)=1$ and $T^{(a,b)}_{1}(u)=
\frac{1}{2}(a+b+2)u+\frac{1}{2}(a-b)$. 

The equivalence between the spin-s spherical harmonics and Jacobi polynomials 
goes as follows:

\beqr
P^{l}_{m,n}(u) &=& N^{l}_{m,n} \left(\frac{1-u}{2}\right)^{\frac{a}{2}}
\left(\frac{1+u}{2}\right)^{\frac{b}{2}}T^{(a,b)}_{l-m}(u) \non\\
&=& N^{l}_{m,n}\left(\frac{1-u}{2}\right)^{\frac{a}{2}}
\left(\frac{1+u}{2}\right)^{\frac{b}{2}}(-1)^{k}T^{(b,a)}_{l-m}(-u) \non\\
&=& N^{l}_{m,n}\frac{(-1)^{l-m}}{2^l}(1-u)^{\frac{a}{2}}
(1+u)^{\frac{b}{2}}\sum_{\nu=0}^{l-m}\left( \begin{array}{c} l+n\\ l-m-\nu 
\end{array} \right) \left( \begin{array}{c} l-n\\ \nu 
\end{array} \right)(-1)^{\nu} \non\\
&\times& \left( \frac{1+u}{1-u} \right) (1-u)^{l-m} \non\\
&=& N^{l}_{m,n}\left( \frac{1-u}{2} \right)^{l} \non\\
&\times& \sum_{\nu=0}^{l-m} \left( \begin{array}{c} l+n\\ \nu+m+n 
\end{array} \right) \left( \begin{array}{c} l-n\\ \nu \end{array} 
\right)(-1)^{l-m+\nu} \left(\frac{1+u}{1-u} \right)^{\nu-(\frac{m+n}{2})}.
\label{spin-s}
\feqr

\noi where for simplicity we assumed that $|m-n|, |m+n|\geq 0$. Setting in 
(\ref{spin-s}) $m=-m$ we obtain the expression:

\beqr
P^{l}_{-m,n}(u) &=& N^{l}_{-m,n}\left( \frac{1-u}{2} \right)^{l} \non\\
&\times& \sum_{\nu=0}^{l+m} \left( \begin{array}{c} l+n\\ \nu-m+n 
\end{array} \right) \left( \begin{array}{c} l+n\\ \nu \end{array} 
\right)(-1)^{l+m+\nu} \left(\frac{1+u}{1-u} \right)^{\nu-(\frac{m-n}{2})} 
\non\\
&=& N^{l}_{-m,n}(\sin \frac{\theta}{2})^{2l} \non\\
&\times& \sum_{\nu=0}^{l+m}\left( \begin{array}{c} l+n\\ \nu-m+n 
\end{array} \right) \left( \begin{array}{c} l-n\\ \nu \end{array} 
\right)(-1)^{l-m-\nu} (\cot \frac{\theta}{2})^{2\nu-m+n}
\label{golb} 
\feqr

\noi which is the one found in \cite{golberg}. The normalization factor 
$N^{l}_{m,n}$ can be constructed by exploring initially the case when 
$a=b=|m|$. Then one finds, making use of the orthogonality of the associated 
Legendre functions $\int_{-1}^{1}[P^{l}_{m}(u)]^2 du
=\frac{2}{2l+1}\frac{(l+m)!}{(l-m)!}$, that:

\beq
\frac{(l!)^2}{[(l+m)!]^2}\int_{0}^{2\pi}d\phi \int_{-1}^{1}du 
P^{l}_{m}(u)P^{l}_{m}(u)=\frac{4\pi}{2l+1}\frac{(l!)^2}{(l+m)!(l-m)!}.
\feq{ortleg}  

\noi Thus $N^{l}_{m,n=0}=\left[\frac{2l+1}{4\pi}\frac{(l-m)!(l+m)!}{(l!)^2} 
\right]^{\frac{1}{2}}$.

%%%%%%%%%%%%%%%%%%%%%%%%%%%%%%%%%%%%%%%%%%%%%%%%%%%%%%%%%%%%%%%%%%%%%%%%%%%%%%%%%%%%%%%%%

\end{document}